\documentclass[12pt,showpacs,twocolumn, preprintnumbers,aps,pre, reprint,superscriptaddress,nofootinbib]{revtex4-1}

\usepackage{graphicx}
\usepackage{subfigure}
\usepackage[labelfont={footnotesize,bf},textfont={footnotesize}]{caption}
\usepackage{color}
\usepackage{amsmath,amssymb,mathtools}
\usepackage{epstopdf}
\usepackage[linktocpage,colorlinks=true,linkcolor=blue,citecolor=blue, urlcolor = black, breaklinks=true]{hyperref}
\usepackage{url}
\usepackage{verbatim}
\usepackage{breakcites}
\usepackage[version=3]{mhchem}

\newcommand{\be}{\begin{equation}}
\newcommand{\ee}{\end{equation}}

\makeatother

\begin{document}

\preprint{}

\title{Analytical solution of stochastic resonance in the non-adiabatic regime}

\author{Woosok Moon }
\affiliation{Department of Mathematics, Stockholm University 106 91 Stockholm, Sweden}
\affiliation{Nordita, Royal Institute of Technology and Stockholm University, SE-10691 Stockholm, Sweden}
\email[]{woosok.moon@gmail.com\\ludovico.giorgini@su.se\\john.wettlaufer@yale.edu}

\author{L. T. Giorgini}
\affiliation{Nordita, Royal Institute of Technology and Stockholm University, SE-10691 Stockholm, Sweden}

\author{J. S. Wettlaufer}
\affiliation{Yale University, New Haven, USA}
\affiliation{Nordita, Royal Institute of Technology and Stockholm University, SE-10691 Stockholm, Sweden}

\date{\today}

\begin{abstract}
We generalize stochastic resonance to the non-adiabatic limit by treating the double-well potential using two quadratic potentials.  
We use a singular perturbation method to determine an approximate analytical solution for the probability density function that asymptotically connects 
local solutions in boundary layers near the two minima with those in the region of the maximum that separates them.  The validity of the analytical solution is 
confirmed numerically.  Free from the constraints of the adiabatic limit, the approach allows us to predict the escape rate from one stable basin to another for systems experiencing a more complex periodic forcing.

\end{abstract}

\pacs{}

\maketitle

\section{Introduction}

It is common to consider noise as a hindrance in measurements and observations, which underlies the general idea of filtering in signal processing \cite{lang1996,chen2006}. In contrast, there are circumstances in which the presence of noise may facilitate the detection of a signal.
A prominent example is {\em stochastic resonance}, during which noise can amplify a weak signal and drive a dramatic transition in the state of a system \cite{benzi1981, benzi1982, benzi1983, mcnamara1989, gammaitoni1998}.  The simplest stochastic resonance configuration considers the trajectory of a Brownian particle
in a double-well potential influenced by a weak periodic forcing; as the random forcing increases so too does the observed signal-to-noise ratio \cite{mcnamara1989}?   As the noise amplitude varies a resonance with the periodic forcing triggers transitions between the stable minima.

The term stochastic resonance originated in a series of studies by Benzi et al., \cite{benzi1981, benzi1982, benzi1983} that focused on the observed periodicity of Earth's ice ages.  Namely, because the 100 kyr eccentricity of the Milankovitch orbital cycles provides such a weak periodic solar insolation forcing, as the strength of random forcing varies, a resonance may drive the transition between the cold and warm states of the Earth's climate system. Whilst the original motivation was an explanation of ice-age periodicity, the generality of the framework has driven a myriad of studies across science and engineering.  For example, a modest subset in which stochastic resonance is a key process includes the sensory systems of many animals, including humans, facilitating the recognition of weak signals buried in environment noise \cite{levin1996,  russell1999, kitajo2003}, it is used to enhance signals in blurry images \cite{rallabandi2010, chouhan2013} and in detecting machine faults in mechanical engineering \cite{lu2019}.  Despite this breadth, the theoretical foundation of stochastic resonance 
is based on the Kramer's escape rate from one of the mimina of a bi-stable system within the adiabatic limit \cite{mcnamara1989, gammaitoni1998}.   In particular, rather than directly solving the non-autonomous Fokker-Planck equation, the periodic forcing of the potential is treated as a constant assuming that the frequency 
of the periodic forcing is asymptotically small. However, in most realistic settings, a weak signal is not consistent with a single periodic function, but rather with a continuous spectrum of many frequencies \cite{collins1996, lu2019, moss2004, mcinnes2008}. However, in other problems, for example when studying the crossover between subcritical and supercritical noise intensity regimes in periodically forced bistable systems \cite{nader2021stochastic, berglund2002sample}, the small periodic forcing assumption must be abandoned.
Therefore, use of stochastic resonance in practical systems requires a generalization of the existing theory to the {\em non-adiabatic case}.  

We have made a recent advance in this direction as follows \cite{moon2020}.  We first treated the Kramers escape problem with periodic forcing within the framework of singular perturbation theory and matched asymptotic expansions.  In particular, we divided the cubic potential into three regions--two boundary layers near the extrema and one connecting them--and determined the local approximate analytic solutions of the Fokker-Planck equation within each region.  We constructed a uniformly valid composite global solution by systematic asymptotic matching of the local solutions.  The bi-stable system was then obtained by reflection of the Kramers problem.  In consequence, the non-adiabatic hopping rate was determined and the full solution was tested numerically.  The analytic solution is principally reliable only when the magnitude of the periodic forcing is much smaller than the noise magnitude.  Our goal here is to simplify this calculation to make more transparent and useful the general treatment of non-adiabatic stochastic resonance.  

Insight for this simplification is provided from a calculation of the mean first passage time of the Ornstein-Uhlenbeck process using similar singular perturbation methods \cite{giorgini2020}.  In that problem we solve the Fokker-Planck equation with an absorbing boundary condition at a point far away from the minimum of a quadratic potential.  In what follows we revisit this problem and derive the probability density function from which we determine the escape rate.  We will then use this method to treat the stochastic resonance problem as a periodically forced bi-stable potential by combining two quadratic potentials.  We then obtain approximate analytical solutions for stochastic resonance in the non-adiabatic regime.  The analytical solutions compare extremely well with the numerical solutions.

\section{First Passage Problem for the Ornstein-Uhlenbeck process}\label{sec:fpp}

\begin{figure}[ht]
\centering
\includegraphics[angle=0,scale=0.22, trim= 0mm 0mm 0mm 0mm, clip]{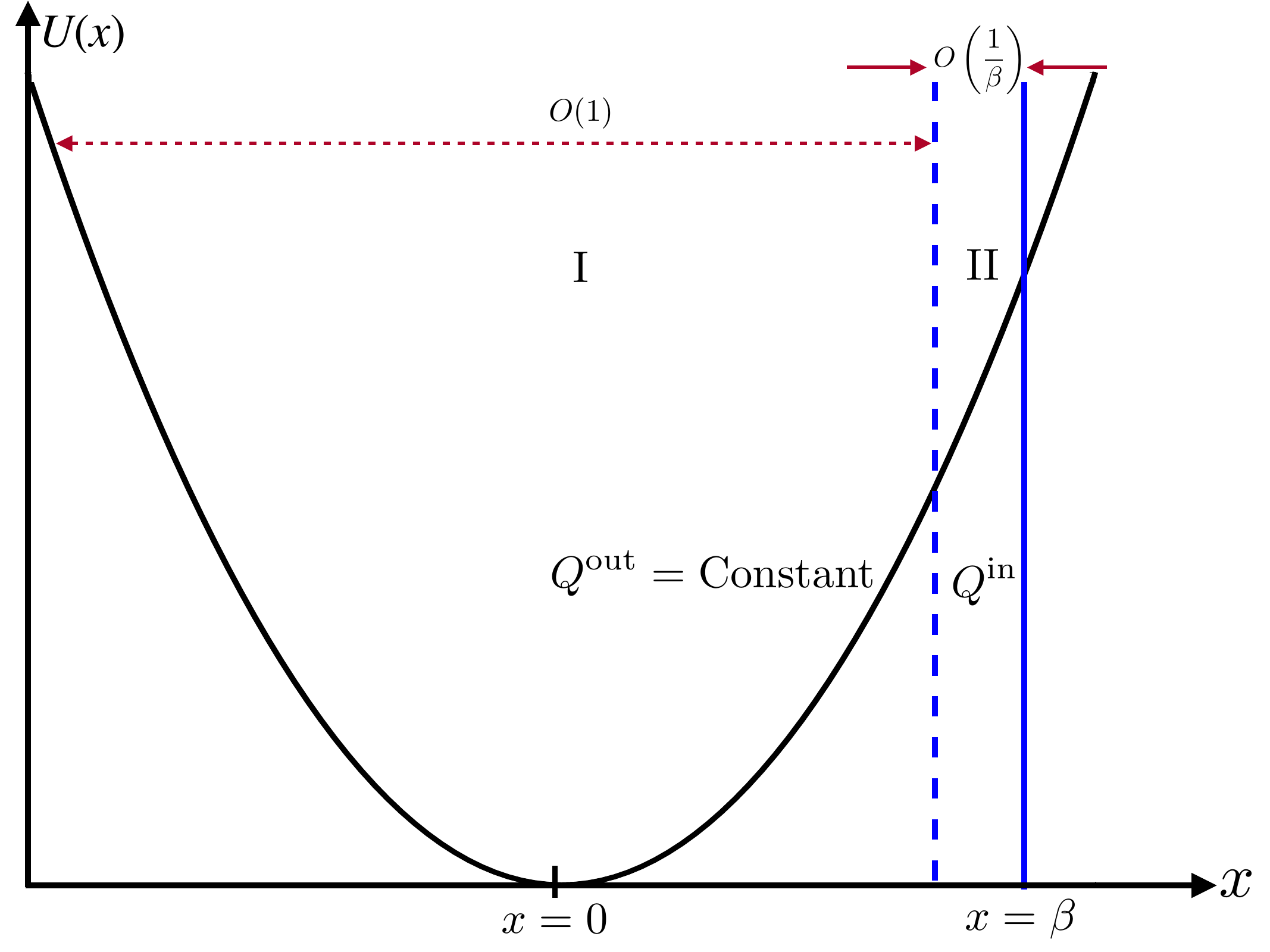}
\caption{Schematic of the first passage problem with a quadratic potential $U(x)=\frac{1}{2}\lambda x^2$ and Gaussian white noise of magnitude $\sqrt{2}\sigma$.
The potential is divided into an ``outer'' region (I) around the minimum at $x=0$, and a ``boundary layer'', or the ``inner'' region (II), near the absorbing boundary at $x=\beta$.  In each region the asymptotically dominant solutions are determined and then matched throughout the entire domain. The boundary layer solution,
$Q^{\text {in}}$, satisfying the boundary condition $x=\beta$ converges to $Q^{\text {out }}$ as the independent variable approaches the origin.  
}
\label{fig:schematic01}
\end{figure}

In order to make this paper reasonably self-contained, we outline the asymptotic method previously used to solve the survival probability problem for the Ornstein-Uhlenbeck problem \cite{giorgini2020}, which we relate explicitly in this section to the first passage problem. 
As shown in Fig. \ref{fig:schematic01} we divide the domain into two regions:  a broad $O(1)$ region ($I$) containing the minimum of the potential, $x=0$, and a narrow $O\left({1}/{\beta}\right)$ boundary layer ($II$) near $x=\beta$.   In the parlance of asymptotic methods in differential equations \cite{BO}, the boundary layer is referred to as the {\em inner region} and the remainder of the domain is the {\em outer region}, although in this case the latter is in the interior of the potential.  We solve the limiting differential equations in these regions, from which we develop a uniform composite solution for the probability density using asymptotic matching.  From this composite solution we calculate the mean first passage time.  

A Brownian particle resides in a quadratic potential (Fig.~\ref{fig:schematic01}) and its position, $x(t)$, obeys the following Langevin equation,
\begin{align}
 \frac{dx}{dt} = -\lambda x + \sqrt{2}\sigma\xi(t),
 \label{eq:lang}
\end{align}
where $\lambda > 0$ captures the stability of the quadratic potential, $\xi(t)$ is Gaussian white noise with zero mean, intensity $\sqrt{2}\sigma$, and 
correlation $\langle\xi(t) \xi(s)\rangle=\delta(t-s)$.

The first passage problem describes the mean time required for the Brownian particle to reach a particular location, say $x=\beta$. 
The Langevin equation can be transformed to a Fokker-Planck equation for the probability density, $P(x,t)$, given by
\begin{align}\label{eq:mainFP}
\frac{\partial P}{\partial t} = \lambda\frac{\partial}{\partial x}(xP)+\sigma^2\frac{\partial^2 P}{\partial x^2},
\end{align}
with an absorbing boundary condition at $x=\beta$, $P(x=\beta)=0$, and $P(x=-\infty) = 0$.
Here, the absorbing boundary condition implies that Brownian particles disappear at $x=\beta$, where we seek the loss rate of probability density.
We have solved this problem analytically for $\beta \gg 1$
with both $\lambda$ and $\sigma = O(1)$ \cite{giorgini2020}.

\underline{ Region I}: {\em The outer solution}.  The probability density in the outer region (the interior of the potential), $P^{\text {out }}$, satisfies
\begin{align}
\label{eq:pout}
\frac{\partial P^{\text {out }}}{\partial t} = \lambda\frac{\partial}{\partial x}\left(x P^{\text {out }}\right)+\sigma^2\frac{\partial^2 P^{\text {out }}}{\partial x^2},
\end{align}
with boundary conditions $P^{\text {out }}(\pm\infty)=0$. The steady-state solution of Eq. (\ref{eq:pout}) is ${P_\text{ss}}^{\text {out }}=\sqrt{\frac{\lambda}{2\pi\sigma^2}}\text{exp}(-\lambda x^2/2\sigma^2 )$.
Now, we let $P(x,t)={P_\text{ss}}^{\text {out }}Q(x,t)$ in Eq. (\ref{eq:mainFP}) and hence $Q(x,t)$ satisfies 
\begin{align} \label{eq:q}
\frac{\partial Q}{\partial t} = -\lambda x\frac{\partial Q}{\partial x}+\sigma^2\frac{\partial^2 Q}{\partial x^2}.
\end{align} 
Therefore, in the interior of the potential $Q \equiv Q^{\text {out }}\approx N(t)$, where $N(t)$ is a very slowly-varying function due to the leaking of probability density at $x=\beta$. Thus, we set $N(t) = N$ to be a constant.

\underline{ Region II}: {\em The inner solution}.  In the boundary layer we have $x \sim \beta$  and because $\beta \gg 1$, we have 
$\varepsilon \equiv 1 / \beta \ll 1$, motivating the stretched coordinate $\eta=\frac{x-\frac{1}{\varepsilon}}{\varepsilon} =\beta(x-\beta)$.
Thus, expressed using $\eta$, in the boundary layer Eq. (\ref{eq:q}) becomes 
\begin{align} \label{eq:qin1}
\frac{\partial Q^{\text {in}}}{\partial t} = 
-\lambda(\eta+\beta^2)\frac{\partial Q^{\text {in}}}{\partial\eta}+\beta^2\sigma^2\frac{\partial^2Q^{\text {in}}}{\partial \eta^2}. 
\end{align}
Because $\beta \gg 1$ the leading-order balance in Eq. (\ref{eq:qin1}) is
\begin{align}
 -\lambda\frac{\partial Q^{\text {in}}}{\partial\eta}+\sigma^2\frac{\partial^2Q^{\text {in}}}{\partial\eta^2} \simeq 0.
\end{align}
Therefore, $Q^{\text {in}} \simeq K_1\text{exp}(\lambda\eta/\sigma^2)+K_2$ with constants $K_1$ and $K_2$. To satisfy the boundary condition $P(x=\beta)=0$, 
which is equivalent to $Q^{\text {in}}(\eta=0)=0$, we must have $K_2+K_1=0$.  

\underline{Uniformly Valid Composite Solution}: Asymptotic matching of the solutions between Regions I and II requires that 
$\lim_{\eta\to -\infty}Q^{\text {in}} = Q^{\text {out }} = N$, which gives $Q^{\text {in}} = N[1-\text{exp}(\lambda\eta/\sigma^2)]$.
A uniformly valid composite solution is then $Q\equiv Q^{\text {out }}+Q^{\text {in}}-\lim_{\eta\to -\infty}Q^{\text {in}}$, or 
$Q\simeq N\left(1-\text{exp}[\beta\lambda(x-\beta)/\sigma^2]\right)$, which gives
\begin{align}
P \simeq N\sqrt{\frac{\lambda}{2\pi\sigma^2}}\text{exp}\left(-\frac{\lambda}{2\sigma^2}x^2\right)\left(1-\text{exp}\left[\frac{\beta\lambda}{\sigma^2}(x-\beta)\right]\right).
\end{align} 
Integrating the Fokker-Planck equation (\ref{eq:mainFP}) over the entire domain $(-\infty, \beta)$ we have 
\begin{align}\label{eq:flux_domain}
\frac{\partial}{\partial t}\int_{-\infty}^{\beta}Pdx = J_{x=\beta},
\end{align}
where $J=\lambda x P+\sigma^2\partial P/\partial x$ and we use $J(x=-\infty) = 0$. Probability density is principally concentrated near $x=0$, and hence
\begin{align}
\int_{-\infty}^{\beta}Pdx \simeq Q^{\text {out }}\int_{-\infty}^{\infty} P^{\text {out }}dx = N, 
\end{align}
so that Eq. (\ref{eq:flux_domain}) becomes
\begin{align}
\frac{dN}{dt} = -\beta\lambda\sqrt{\frac{\lambda}{2\pi\sigma^2}}\text{exp}\left(-\frac{\lambda\beta^2}{2\sigma^2}\right)N\equiv-rN.
\end{align}
Therefore, the global probability density decreases with rate $r$ and the mean first passage time is
$\langle T\rangle = 1/r$, with
\begin{align}
r=\beta\lambda\sqrt{\frac{\lambda}{2\pi\sigma^2}}\text{exp}\left(-\frac{\lambda\beta^2}{2\sigma^2}\right).
\end{align}

\section{Stochastic resonance in double quadratic potentials}\label{sec:sto_res}

\begin{figure}[ht]
\centering
\includegraphics[angle=0,scale=0.22,trim= 0mm 0mm 0mm 0mm, clip]{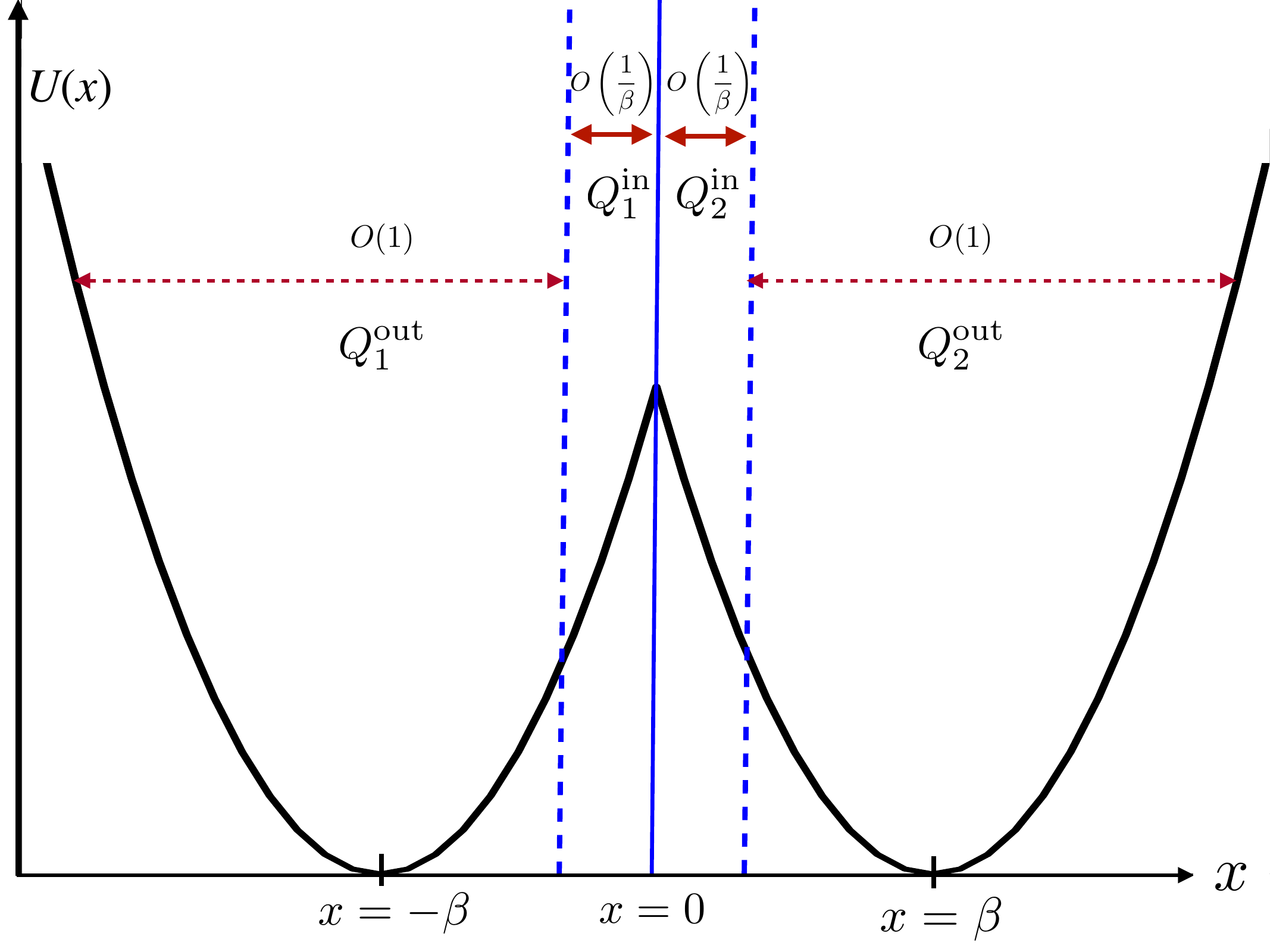}
\caption{Schematic of stochastic resonance in which a double-well potential is treated using two quadratic potentials, 
$U_1(x)=\frac{1}{2}\lambda (x+\beta)^2$ when $x<0$ and $U_2(x)=\frac{1}{2}\lambda (x-\beta)^2$ when $x \ge 0$. 
There are two regions, the interior of each near the two minima $x=\pm\beta$ and boundary layers near $x=0$.  }
\label{fig:schematic02}
\end{figure}

\subsection{Asymptotic solutions}

We now combine the asymptotic methods used to solve the general problem of stochastic resonance \cite{moon2020} with the particular setting 
described in \S \ref{sec:fpp}, to treat the double-well potential in stochastic resonance using two quadratic potentials. 
As shown in Fig.~\ref{fig:schematic02}, the potential $U(x)$ is $U_1(x) = \frac{1}{2}\lambda(x+\beta)^2$ when $x<0$ 
and $U_2(x)=\frac{1}{2}\lambda(x-\beta)^2$ when $x\ge 0$. Thus, with the addition of noise, $\sqrt{2}\sigma\xi(t)$, and periodic forcing $A\text{cos}(\omega t)$, stochastic resonance between the minima at $x=\pm\beta$ is described by the following Fokker-Planck equation for the probability density $P(x,t)$, 
\begin{align}\label{eq:interior}
\frac{\partial P}{\partial t} = \frac{\partial}{\partial x}\left\{\left[\frac{dU}{dx}-A\text{cos}(\omega t)\right]P\right\}+\sigma^2\frac{\partial^2 P}{\partial x^2}, 
\end{align}
with boundary conditions  $P(x=\pm\infty,t) = 0$. Here, the magnitude of the periodic forcing $A$, $\lambda$ and $\sigma$ are all 
order $O(1)$ quantities.

As shown in the Fig.~(\ref{fig:schematic02}), there are two {\em outer} regions interior to each side of the potential centered upon 
the two minima, $x=\pm\beta$, and two {\em boundary layers} for each quadratic potential at $x=0$. We now consider approximate solutions to Eq.~(\ref{eq:interior}) in these regions. 

In the outer region of $U_1(x)$ centered on $x=-\beta$, the probability density $P^{\text {out }}_1$ satisfies
\begin{align}\label{eq:out1}
\frac{\partial P^{\text {out }}_1}{\partial t}=\lambda\frac{\partial}{\partial y}(yP^{\text {out }}_1)
-A\text{cos}(\omega t)\frac{\partial P^{\text {out }}_1}{\partial y}+\sigma^2\frac{\partial^2P^{\text {out }}_1}{\partial y^2}
\end{align}
with $P^{\text {out }}_1(y=\pm\infty)=0$, where $y=x+\beta$. Eq.~(\ref{eq:out1}) has solution
\begin{align}
P^{\text {out }}_1=\sqrt{\frac{\lambda}{2\pi\sigma^2}}\text{exp}\left(-\frac{\lambda}{2\sigma^2}[y-h(t)]^2\right)
\end{align}
where
\begin{align}
h(t)=\frac{A}{\sqrt{\lambda^2+\omega^2}}\text{cos}(\omega t - \phi)
\end{align}
and $\text{tan}\phi = \omega/\lambda$.

Now we substitute $P(x,t)=P^{\text {out }}_1Q(x,t)$ into Eq.~(\ref{eq:interior}) which becomes
\begin{align}
\frac{\partial Q}{\partial t}=[-\lambda y+A\text{cos}(\omega t - 2\phi)]\frac{\partial Q}{\partial y}+\sigma^2\frac{\partial^2 Q}{\partial y^2}, 
\end{align}
the outer solution of which is $Q^{\text {out }}_1=N_1(t)$; a slowly-varying function of time that we approximate as a constant, $N_1$.

In the boundary layer we introduce the stretched coordinate $\eta=\beta x$, which leads to
\begin{align} \label{eq:qin}
\frac{\partial Q^{\text {in}}_1}{\partial t} &=[-\lambda\eta-\beta^2\lambda+\beta A\text{cos}(\omega t - 2\phi)]\frac{\partial Q^{\text {in}}_1}{\partial\eta} \nonumber \\
&+\beta^2\sigma^2\frac{\partial^2 Q^{\text {in}}_1}{\partial\eta^2}. 
\end{align} 
Because $\beta \gg 1$, keeping terms to $O(\beta)$ and higher, gives
\begin{align}
\left[-\lambda+\frac{A}{\beta}\text{cos}(\omega t - 2\phi)\right]\frac{\partial Q^{\text {in}}_1}
{\partial\eta}+\sigma^2\frac{\partial^2 Q^{\text {in}}_1}{\partial\eta^2} \simeq 0,
\end{align}
the solution of which is
\begin{align} \label{eq:qin_sol}
Q^{\text {in}}_1=K_1\text{exp}\left(\frac{\lambda-\frac{A}{\beta}\text{cos}(\omega t-2\phi)}{\sigma^2}\eta\right)+K_2, 
\end{align}
with constants $K_1$ and $K_2$ to be determined\footnote{We note that Eq. (\ref{eq:qin}) could be solved using a regular perturbation method by setting
$Q^{\text{in}}_1 \simeq Q^{\text{in}}_{10}+\frac{1}{\beta}Q^{\text{in}}_{11}$, which results in
$Q^{\text{in}}_1 \simeq K_1\text{exp}\left(\frac{\lambda}{\sigma^2}\eta\right)\left(1-\frac{A}{\beta\sigma^2}\text{cos}(\omega t-2\phi)\eta\right)+K_2$.
 However, as noted, to $O(\beta)$, the solution (\ref{eq:qin_sol}) uses $\text{exp}\left(-\frac{A}{\beta\sigma^2}\text{cos}(\omega t - 2\phi)\eta\right)$
 instead of $1-\frac{A}{\beta\sigma^2}\text{cos}(\omega t-2\phi)\eta$, which simplifies the subsequent development at that order.}.
Asymptotic matching requires that the outer limit of the inner solution equal the inner limit of the outer solution; 
$\lim_{\eta\to-\infty}Q^{\text {in}}_1=\lim_{y\to\beta}Q^{\text {out }}_1=N_1$ and hence $K_2 = N_1$.  
Therefore, a uniformly valid composite asymptotic solution in $U_1(x)$ is $Q_1 = Q^{\text {out }}_1 + Q^{\text {in}}_1 - \lim_{\eta\to-\infty}Q^{\text {in}}_1$, giving 
\begin{align}
P_1&=\sqrt{\frac{\lambda}{2\pi\sigma^2}}\text{exp}\left(-\frac{\lambda}{2\sigma^2}[x+\beta-h(t)]^2\right) \nonumber \\
     &\times\left\{N_1+K_1\text{exp}\left(\frac{\beta\lambda-A\text{cos}(\omega t-2\phi)}{\sigma^2}x\right)\right\}.
\end{align}

Similarly, we obtain the solution $P_2$ in $U_2(x)$ as
\begin{align}
P_2&=\sqrt{\frac{\lambda}{2\pi\sigma^2}}\text{exp}\left(-\frac{\lambda}{2\sigma^2}[x-\beta-h(t)]^2\right) \nonumber \\
     &\times\left\{N_2+D_1\text{exp}\left(-\frac{\beta\lambda+A\text{cos}(\omega t-2\phi)}{\sigma^2}x\right)\right\}.
\end{align}

Now we determine the constants $K_1$ and $D_1$ from the continuity of probability density $P(x)$ and flux $J(x)$ at $x=0$,
where $J(x) = [dU/dx - A\text{cos}(\omega t)]P+\sigma^2\partial P/\partial x$.
Firstly, continuity of $P(x)$ at $x=0$ is $P_1(x=0)=P_2(x=0)$, which results in
\begin{align}\label{eq:continuityP}
D_1 = (N_1+K_1)\text{exp}\left(\frac{2\beta\lambda}{\sigma^2}h(t)\right)-N_2.
\end{align}
The fluxes at the origin from both sides are
\begin{align} \label{eq:flux}
&J_1(x=0) =\left[(\beta\lambda+\lambda h(t)-2A\text{cos}(\phi)\text{cos}(\omega t - \phi))K_1 \right. \nonumber \\
            & \left. +(\lambda h(t)-A\text{cos}(\omega t))N_1\right]
            \sqrt{\frac{\lambda}{2\pi\sigma^2}}\text{exp}\left(-\frac{\lambda}{2\sigma^2}[\beta-h(t)]^2\right)  \nonumber \\
             \text{and}\nonumber \\
&J_2(x=0) =\left[(-\beta\lambda+\lambda h(t)-2A\text{cos}(\phi)\text{cos}(\omega t - \phi))D_1 \right. \nonumber \\
            & \left. +(\lambda h(t)-A\text{cos}(\omega t))N_2\right]
            \sqrt{\frac{\lambda}{2\pi\sigma^2}}\text{exp}\left(-\frac{\lambda}{2\sigma^2}[\beta+h(t)]^2\right),           
\end{align}
respectively.  
Now, imposing $J_1(x=0) =J_2(x=0)$ and $P_1(x=0)=P_2(x=0)$, viz., Eq.~(\ref{eq:continuityP}), gives 
\begin{align} \label{eq:k1d1}
&K_1 = \left[\frac{1}{2}+\frac{A}{2\beta\lambda}\text{cos}(\omega t - 2\phi)\right] \nonumber \\
&\times\left[-N_1+N_2\text{exp}\left(-\frac{2\beta\lambda}{\sigma^2}h(t)\right)\right] \nonumber \\
 &D_1 = \left[\frac{1}{2}-\frac{A}{2\beta\lambda}\text{cos}(\omega t - 2\phi)\right] \nonumber \\
 &\times \left[-N_2+N_1\text{exp}\left(\frac{2\beta\lambda}{\sigma^2}h(t)\right)\right].
\end{align}
Integration of Eq.~ (\ref{eq:interior}) from $x=-\infty$ to $x=0$,
\begin{align}
\frac{\partial}{\partial t}\int_{-\infty}^{0}P_1dx = J_1|_{x=0},
\end{align} 
gives
\begin{align}\label{eq:N1}
\frac{dN_1}{dt} = -r_1N_1+r_2N_2, 
\end{align}
with escape rates
\begin{align} \label{eq:rate}
&r_1=\left[\frac{1}{2}-\frac{A}{2\beta\lambda}\text{cos}(\omega t - 2\phi)\right] [\beta\lambda-\lambda h(t)+A\text{cos}(\omega t)  \nonumber \\
&+A\text{cos}(\omega t - 2\phi)] \sqrt{\frac{\lambda}{2\pi\sigma^2}}
\text{exp}\left(-\frac{\lambda}{2\sigma^2}[\beta-h(t)]^2\right) \nonumber \\
  \text{and}\nonumber \\
&r_2=\left[\frac{1}{2}+\frac{A}{2\beta\lambda}\text{cos}(\omega t - 2\phi)\right][\beta\lambda+\lambda h(t)-A\text{cos}(\omega t)  \nonumber \\
&-A\text{cos}(\omega t - 2\phi)]\sqrt{\frac{\lambda}{2\pi\sigma^2}}
\text{exp}\left(-\frac{\lambda}{2\sigma^2}[\beta+h(t)]^2\right),
\end{align}
where, because $\int_{-\infty}^{\infty}Pdx = 1$, we use the normalization condition $N_1+N_2=1$. 
Similarly, upon integration of Eq.~ (\ref{eq:interior}) from $x=0$ to $x=\infty$, we obtain
\begin{align}\label{eq:N2}
\frac{dN_2}{dt} = r_1N_1-r_2N_2.
\end{align}
Therefore, $r_1$ and $r_2$ are the escape rates from $x=\pm\beta$ over the barrier at the origin and  Eqs.~(\ref{eq:N1}) and (\ref{eq:N2}) are simplified forms 
of the two-state Master equations derived for general quartic potentials \cite{moon2020}.

\subsection{The validity of asymptotic solutions}

\begin{figure}[ht]
\centering
\includegraphics[angle=0,scale=0.24,trim= 8mm 0mm 8mm 8mm, clip]{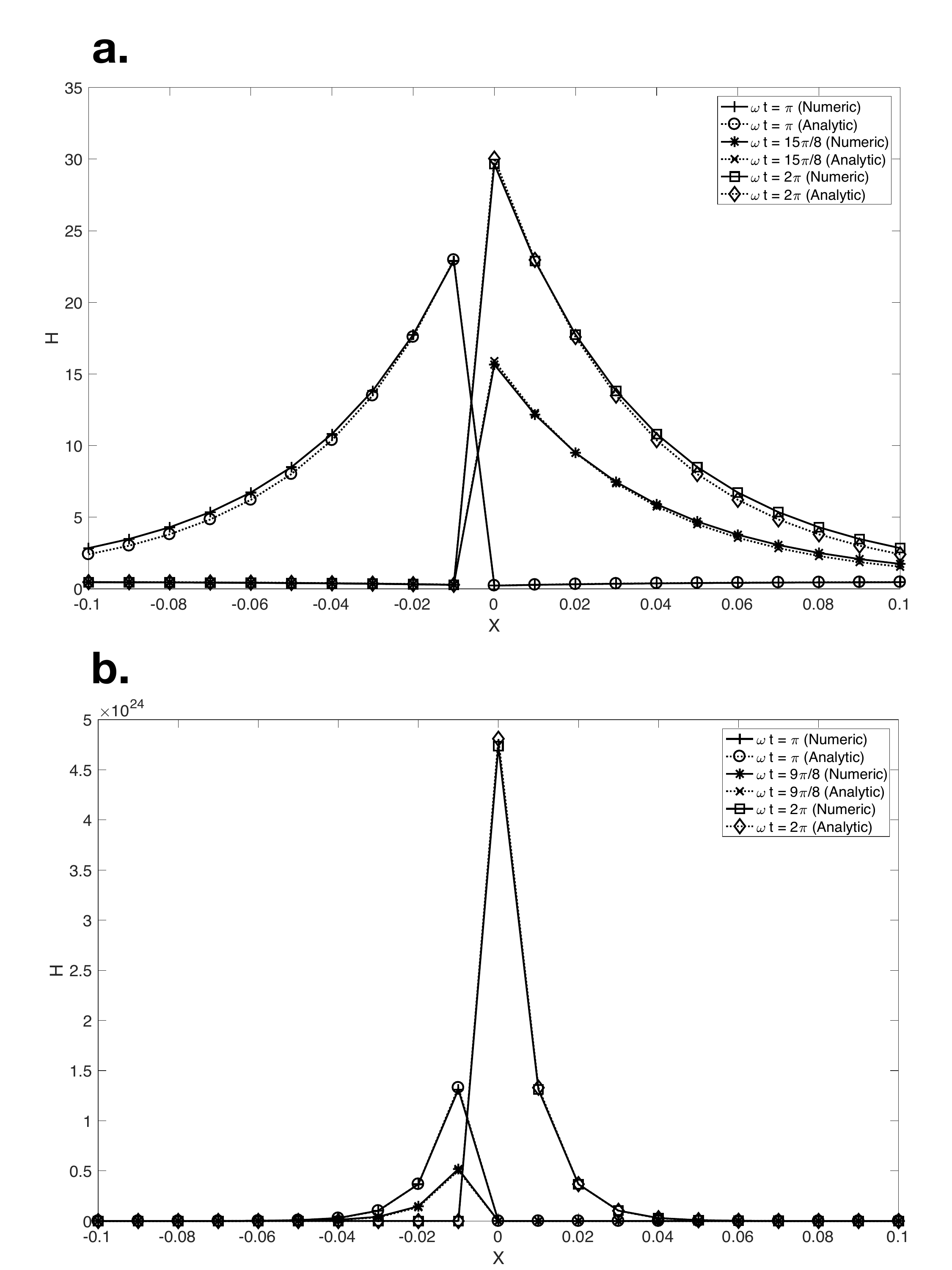}
\caption{The comparison between numerical simulations and asymptotic solutions
using $H(x,t) = P_1(x,t)/P^{\text {out }}_1$ when $x<0$ and $H(x,t)=P_2(x,t)/P^{\text {out }}_2$ when $x \ge 0$.
The different cases are denoted in the inset.  
The two examples use 
$\omega=\pi/20$,  $\beta=1.0$, and $\lambda=1.0$ as common variables. The first case (a) has $A=0.1$ and $\sigma=0.2$ 
and the second case (b) has $A=0.3$ and $\sigma=0.1$.}  
\label{fig:fig03}
\end{figure}

We test the validity of the asymptotic solutions by comparison with full numerical solutions 
of the Fokker-Planck equation (\ref{eq:interior}) using the implicit finite difference method of Chang and Cooper \cite{chang1970}. 
To facilitate this comparison, near $x=0$ we introduce a function defined as $H(x)=P_1/P^{\text {out }}_1$ when $x<0$ and
$H(x)=P_2/P^{\text {out }}_2$ when $x \ge 0$. We study two different magnitudes of the 
periodic forcing, $A$, and the noise magnitude, $\sigma$, and combined with the range of the angular frequency $\omega$, this puts our results in the non-adiabatic regime. In particular, $A$ is not asymptotically smaller than $\sigma$, and $\omega$ is not trivially small.   Previously we showed that the adiabatic limit requires $\omega \ll 1$ and $A \ll \sigma$ and the non-adiabatic case referred to non-trivial $\omega$ \cite{moon2020}.  
Here, we extend the non-adiabaticity with the condition $A \sim \sigma$.  Figure \ref{fig:fig03} shows that the asymptotic and numerical solutions compare very well. When $A \sim \sigma$ or $A\ll1$ (not shown), the asymptotic solutions are nearly indistinguishable from numerical solutions. However, when $A\gg\sigma$, the analytic solutions become less accurate.

\begin{figure}[ht]
\centering
\includegraphics[angle=0,scale=0.23,trim= 0mm 0mm 0mm 0mm, clip]{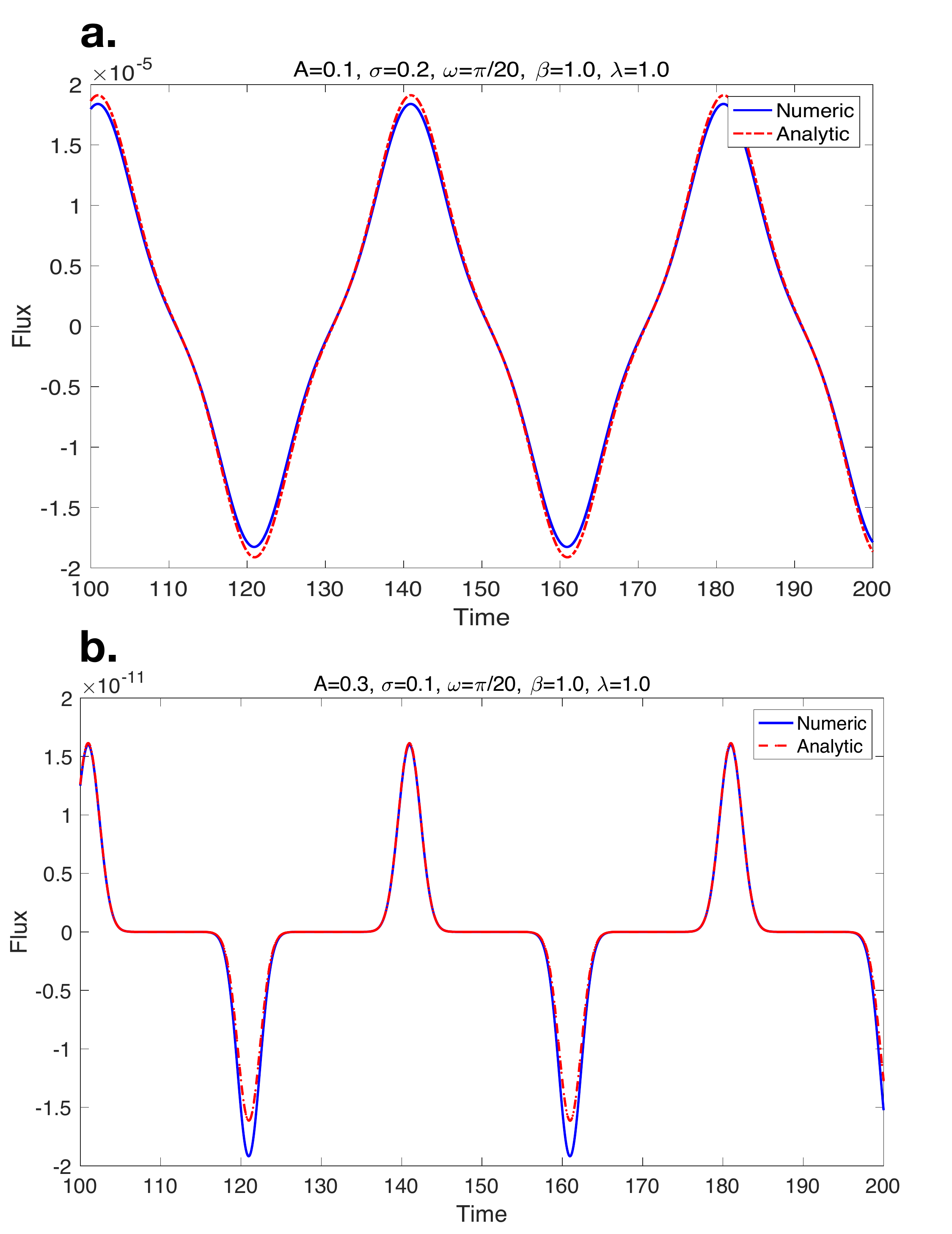}
\caption{The probability flux calculated at the barrier ($x=0$) between the two wells of the potential.  Comparison between numerical simulations (solid lines) 
and asymptotic solutions (dashed lines from Eq.~\ref{eq:flux}) for the same two cases used in Fig. \ref{fig:fig03}.}
\label{fig:fig04}
\end{figure}

The central quantity in stochastic resonance is the flux at the barrier ($x=0$) between the two wells of the potential, which controls the oscillatory behavior of the probability density. In Figure \ref{fig:fig04} we compare the analytical (dashed lines from Eq.~\ref{eq:flux}) and numerical (solid lines) fluxes at $x=0$.  Clearly these compare very favorably. 

The escape (or hopping) rates $r_1$ and $r_2$ shown in Eq. (\ref{eq:rate}) are valid independent of the magnitude of $\omega$ and $A$, and so long as $A$ is not much larger than $\sigma$, the analytic solutions are very accurate. Therefore, our analytic solutions can be used for the wide range of applications of stochastic resonance that appear in science and engineering, which removes the need for substantial simulations of either the Langevin or Fokker-Planck equations.  

\subsection{Weak periodic forcing $A \ll 1$}

In the original treatment of stochastic resonance \cite{benzi1981, benzi1982, benzi1983}, one has $A\ll1$, which implies that there is a weak signal in a noise dominated background.  Here, we still seek to understand the amplification of the periodic forcing $A\text{cos}(\omega t)$ for $A\ll1$, however 
we  treat $\omega$ as arbitrary so that we are have the non-adiabatic case, which is not within the corpus of the original work.  

Because we are in possession of the probability density function $P(x,t)$,  we can calculate the mean position of a Brownian 
particle as
\begin{align}
\langle x \rangle = \int_{-\infty}^{0}xP_1dx+\int_{0}^{\infty}xP_2dx.
\label{response}
\end{align}
We note that $P_1$ and $P_2$ are principally concentrated near $x=-\beta$ and $x=\beta$, so that 
\begin{align}
&\int_{-\infty}^{0}xP_1dx \nonumber \\
&\simeq \sqrt{\frac{\lambda}{2\pi\sigma^2}}N_1
\int_{-\infty}^{\infty}x\text{exp}\left(-\frac{\lambda}{2\sigma^2}[x+\beta-h(t)]^2\right)dx \nonumber \\
&\simeq -N_1\beta.
\end{align}
Similarly, because $\int_{0}^{\infty}xP_2dx \simeq N_2\beta$, we have $\langle x\rangle = \beta(-N_1+N_2)$.
Using Eqs. (\ref{eq:N1}) and (\ref{eq:N2}), we write the time-evolution of $-N_1+N_2$ as
\begin{align} \label{eq:delN}
 \frac{d}{dt}(-N_1+N_2) = -(r_1+r_2)(-N_1+N_2)+(r_1-r_2).
\end{align}
We now take $A \ll 1$ to find
\begin{align}
 &r_1+r_2 \equiv r \simeq \beta\lambda \sqrt{\frac{\lambda}{2\pi\sigma^2}}\text{exp}\left(-\frac{\lambda\beta^2}{2\sigma^2}\right) ~\text{and} \nonumber \\
 &r_1-r_2 \simeq r\frac{\beta\lambda h(t)}{\sigma^2}, 
\label{eq:rapprox}
\end{align}
and hence the approximate solution of Eq.~ (\ref{eq:delN}) is
\begin{align} \label{eq:meanx}
-N_1+N_2 \simeq \frac{\beta A}{\sigma^2}\text{cos}\phi\text{cos}\psi\text{cos}(\omega t-\phi-\psi),
\end{align}
where $\text{cos}\phi = \lambda/\sqrt{\lambda^2+\omega^2}$ and $\text{cos}\psi = r/\sqrt{r^2+\omega^2}$. Hence,
\begin{align} \label{eq:sr_mag}
\langle x\rangle = \beta(-N_1+N_2) = \beta^2 \frac{A}{\sigma^2}\text{cos}\phi\text{cos}\psi\text{cos}(\omega t-\phi-\psi).
\end{align}
Therefore, the original signal is $A\text{cos}(\omega t)$ and the response, $\langle x\rangle$, is order O($\beta^2$).

\begin{figure}[ht]
\centering
\includegraphics[angle=0,scale=0.23,trim= 0mm 0mm 0mm 0mm, clip]{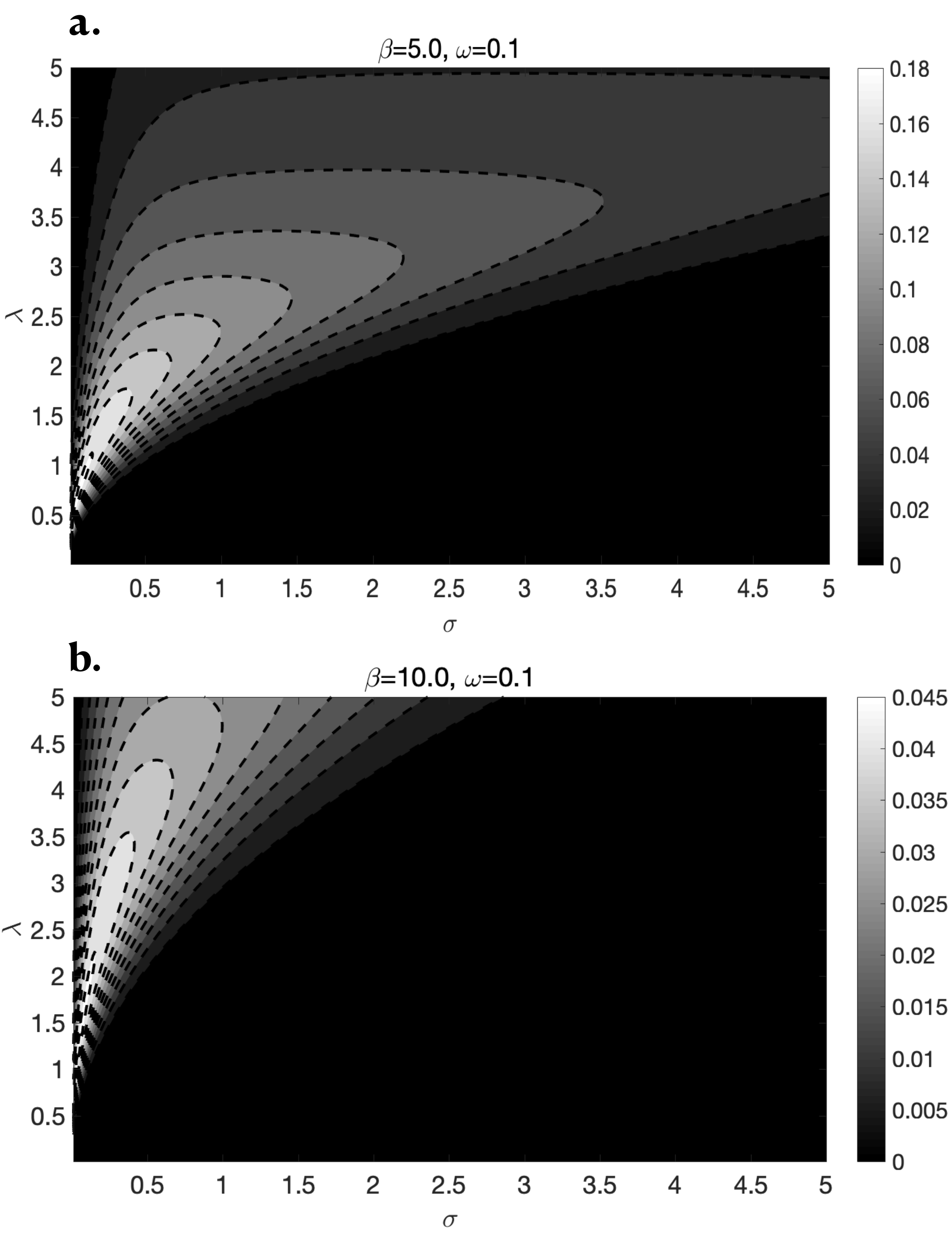}
\caption{Contour plot of the  $\sigma$ and $\lambda$ dependence of the magnitude of the response $\frac{\lambda A}{\sigma^2}\text{cos}\psi$ from Eq.~(\ref{eq:sr_mag}).  The location of the absorbing boundary is (a) $\beta=5$ and (b) $\beta=10$ for a fixed angular frequency of $\omega=0.1$.}
\label{fig:fig05}
\end{figure}

Eq.~(\ref{eq:sr_mag}) shows that the magnitude of the response, $\langle x\rangle$, depends nonlinearly on the stability factor, $\lambda$, 
and the noise amplitude, $\sigma$, as $\frac{\lambda A}{\sigma^2}\text{cos}\psi$, where $\text{cos}\psi$ is a function of $\lambda$ and $\sigma$ (and other parameters) as seen in Eq.~\eqref{eq:rapprox}.
Figure \ref{fig:fig05} shows two contour plots
of the magnitude of the response, $\frac{\lambda A}{\sigma^2}\text{cos}\psi$ for two values of the location of the absorbing boundary (a) $\beta=5$ and (b) 
$\beta=10$ for fixed $\omega=0.1$.  Optimal values of  the noise amplitude, $\sigma$, and the stability of the potential, $\lambda$, are revealed as the maxima in these plots.  This optimal noise magnitude is signature of stochastic resonance.

When $A\ll1$ our results are similar to the adiabatic limit, except for the fact that there is a phase shift, $\phi$.  However, as shown 
above and expected from the original theory, the response to the periodic forcing,  $A\text{cos}(\omega t)$, is magnified by a factor
 $\beta^2$.  In strong contrast to the original theory \cite{mcnamara1989}, we provide an explicit mathematical
expression, Eq. (\ref{eq:sr_mag}), that shows this dependence.  Although we have focused upon a single forcing frequency, our
method can be generalized to weak signals consisting of many frequencies. 

\section{Discussion}

\subsection{Singular perturbation theory and non-adiabaticity}

Here we have studied the Fokker-Planck equation for stochastic resonance directly using singular perturbation theory \cite{BO}, which has been principally developed in applied mathematics and fluid dynamics. The value of this approach is that it overcomes constraints intrinsic to the theory.  The most severe and commonly imposed constraint is adiabaticity, which in this context relies on the direct translation of the Kramers transition rate between the two minima of the double well potential as described in detail presently.   

In adiabatic stochastic resonance the periodic forcing, $A\text{cos}(\omega t)$, is treated under the assumption that $\omega$ is asymptotically small. Hence, the modulation of the potential, $U(x)-A\text{cos}(\omega t)x$, is weak and the Kramers escape rate is correspondingly assumed to be unchanged.  Near a potential minimum, 
$x=x_{min}$,  the dominant time-scale in the dynamics of a Brownian particle is the response time governed by $\lambda=\partial^2U/\partial x^2|_{x=x_{min}}$.  Therefore, in the adiabatic limit $\omega \ll \lambda$, which is heuristically reasonable but poses quantitative challenges, particularly in that there is
no obvious upper bound on $\omega$ to control experimental errors?


The singular perturbation method used here overcomes this limitation of the adiabatic assumption. Rather than assuming the Kramers escape rate is unchanged, 
we construct a time-dependent solution directly without the small $\omega$ assumptions;  $\omega \ll r$ and $\omega \ll \lambda$.  Indeed, 
Eq. (\ref{eq:sr_mag}) is a generalization of the adiabatic theory as reflected through the presence of the extra term $\text{cos}\phi$.  Hence, 
the adiabatic limit is recovered when $\text{cos}\phi \simeq 1$, which is equivalent to $\omega \ll \lambda$, thereby providing a clear quantification of the adiabatic limit and a framework to generalize stochastic resonance to a wide class of non-adiabatic cases and forcing signals.


We note here the other methods that have been introduced to overcome the adiabatic limit; Floquet and linear response theory.  In the former, 
one uses the Floquet theorem to construct a series solution of periodic Floquet modes \cite{jung1993, gang1990}.  This approach does not require 
appeal to a small parameter, but seeks the exact solution using an infinite sum eigenfunctions. However, because the solutions are
a complicated series of eigenfunctions, they can be challenging to use various applications.  
Linear response theory only requires weak, or low amplitude, $A \ll 1$, signals amplified by an undisturbed linear operator \cite[e.g.,][]{dykman1995, anishchenko1999}. 
It can thereby treat non-adiabatic cases, but when $A$ is $O(1)$, the approach clearly has limitations.  
Note that the hopping
rates $r_1$ and $r_2$ in Eq. (\ref{eq:rate}) do not require the condition $A \ll 1$.


We summarize this section by emphasizing that the power of singular perturbation and asymptotic methods is the great simplifications of 
complex differential equations that they provide, whilst giving rise to analytical solutions.  Because the edifice of the approach has been developed
in applied mathematics and fluid dynamics \cite{BO}, where many of the key features of the equations of motion are present in the Fokker-Planck equations appearing in  stochastic problems, we argue that the theoretical tools of asymptotic methods will find a fertile ground in stochastic dynamics.  

 
\subsection{Relationship to a two-state model}

When the hopping time is much longer than the response time of a Brownian particle a minima, it may not be necessary 
to consider the detailed motion of the particle in the vicinity of a minimum.  In this case, a two-state model focusing solely 
on the hopping dynamics provides an approximation of the full solution of the Fokker-Planck equation for stochastic resonance \cite{mcnamara1989}. 
Here we show a clear link between the complete problem and the two-state model.  Namely, for $\beta \gg 1$, we approximate the probability density function, $P(x,t)$, 
by connecting two Gaussians centered around the two minima:  We impose the continuity of $P(x,t)$ and the probability flux $J(x,t)$ to determine the magnitudes of the two Gaussians, $N_1$ and $N_2$.  Thus, because Eqs. (\ref{eq:N1}) and (\ref{eq:N2}) are exactly same as the two-state model with the new hopping rates $r_1$
and $r_2$, we show the equivalence between the complete Fokker-Planck equation with a double-well potential and a two-state model with two hopping rates.


\subsection{Applications of stochastic resonance using a double-quadratic potential}

The canonical approach to stochastic resonance has a continuous quartic potential because it relies on the Kramers
exit rate, which requires the curvature of potential at the minima and the maximum.  Here, as described in \S \ref{sec:fpp}, rather than relying on 
the Kramers exit rate, we treated the first passage problem of an Ornstein-Uhlenbeck process for a double-well potential consisting of two
connected quadratic potentials.  Thus, for a dwell time in either minimum that is considerable, one can consider independent Ornstein-Uhlenbeck
processes in each potential.   This provides a framework for experimental or observational data for which there are insufficient transitions
to deduce the precise functional form of the potential connecting the two minima.  Here we describe the utility of our approach

\subsubsection{How general is the double-quadratic potential?}

\begin{figure}[ht]
\centering
\includegraphics[width=.9\linewidth]{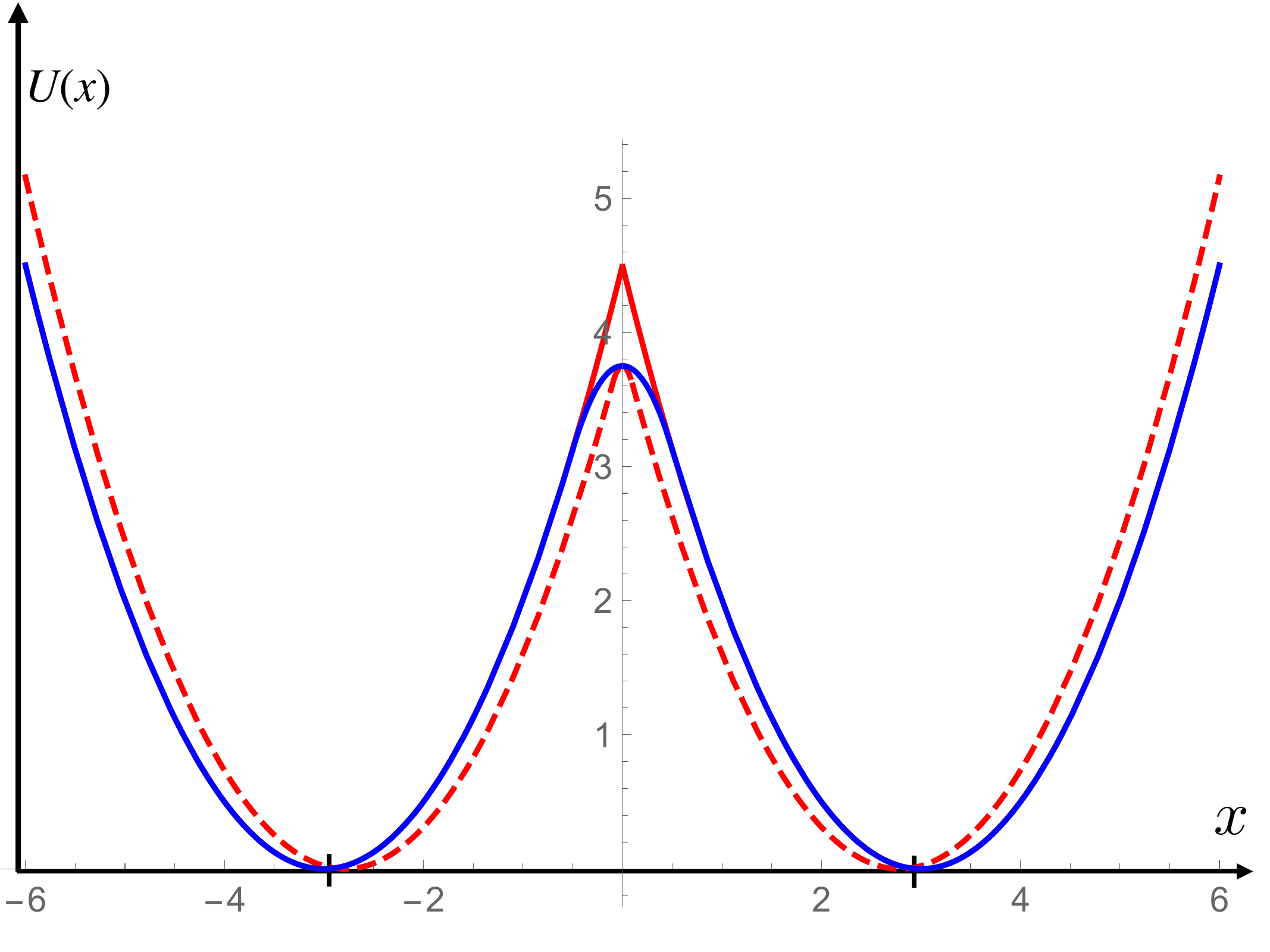}
\caption{Schematic of the potential $U(x)$ defined in Eq. (\ref{potential_U}) (solid blue line) with $\beta=3,\, \delta=0.5$ and of two double parabolic potentials with $\beta=3$ (dashed red line) and $\beta=\sqrt{7.5}$ (solid red line). In all three potentials $\lambda=1$.} 
\label{fig:fig01dis}
\end{figure}

The general question concerning the applicability of our approach is the degree to which our potential structure captures the behavior of a continuous potential with two minima separated by one maximum.  To that end we compare the response in the case of the double parabolic potential, Eq. (\ref{response}), to that in which the potential is approximated using three parabolas.  Thus, we define the following potential by imposing continuity of the three parabolas and of their first derivatives at $x=\pm \delta$  as
\begin{equation}
U(x)=
\begin{cases}
\frac{\lambda}{2}(x+\beta)^2\;\;\; \textrm{for}\; x<-\delta,\\
-\frac{\lambda(-\delta+\beta)}{2\delta}x^2+\frac{\lambda}{2}(-\delta+\beta)\beta\;\;\; \textrm{for}\; x\in[-\delta,\delta],\\
\frac{\lambda}{2}(x-\beta)^2\;\;\; \textrm{for}\; x>\delta.
\end{cases}
\label{potential_U}
\end{equation}
In Fig. (\ref{fig:fig01dis}) we see that the potential in Eq. \eqref{potential_U}
is enveloped by the two double parabolic potentials $U_{1(2)}(x) = \frac{\lambda}{2}(x\pm \beta_{1(2)})^2$ with $\beta_1 = \beta$ and $\beta_2=\sqrt{\beta(\beta-\delta)}$ for $x \in (-\beta_2,\beta_2)$.  We thus choose small but finite values of $\delta$ in Eq. \eqref{potential_U} to examine the response of the general case, where the gradient of the potential vanishes at the origin.  Because by setting $\delta=0$ we recover the double parabolic potential, our aim is to show how our results in that case can be generalized to the case with finite $\delta$. 

\begin{figure}[ht]
\centering
\includegraphics[angle=0,scale=0.53,trim= 30mm 0mm 0mm 0mm, clip]{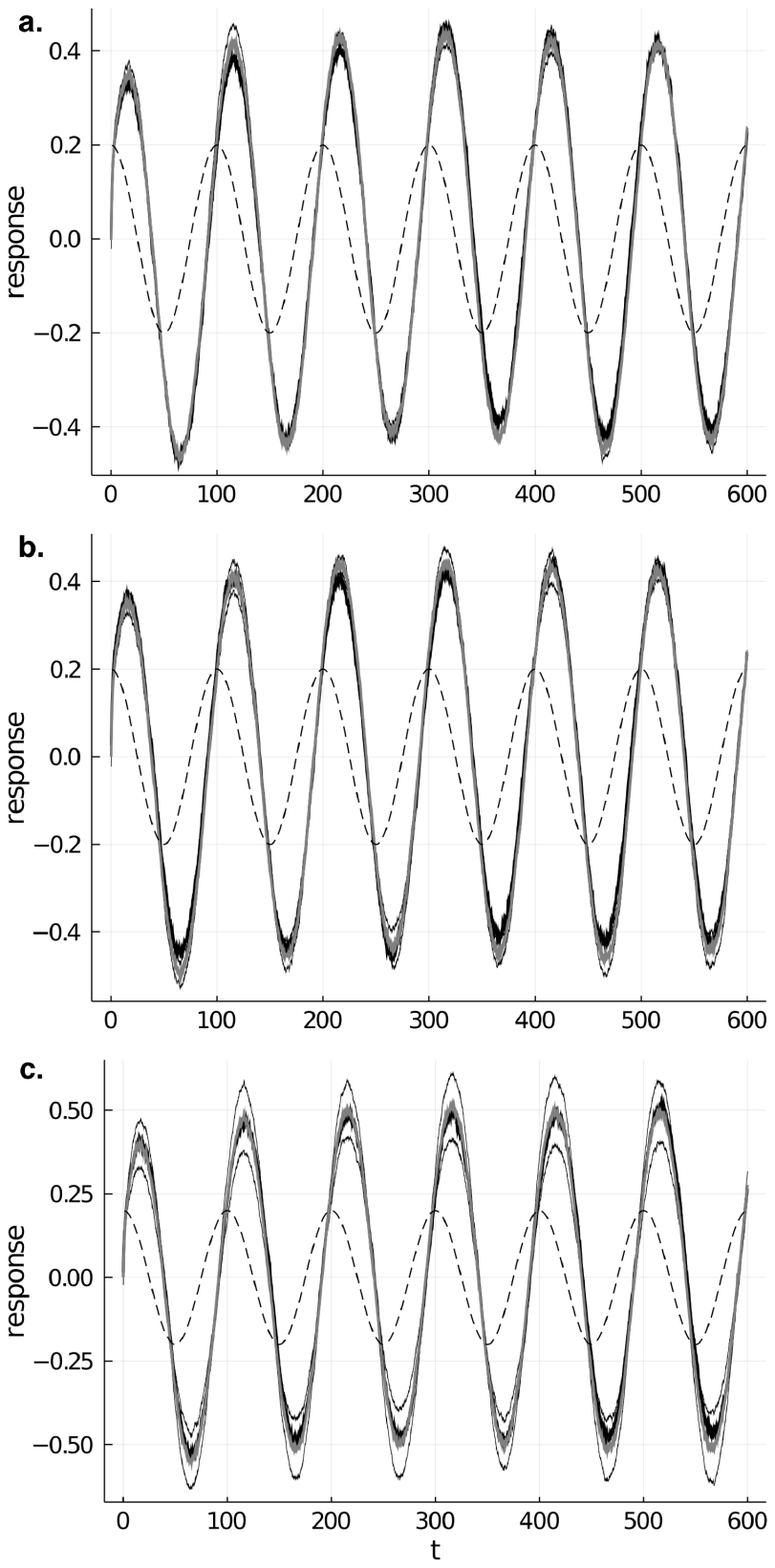}
\caption{Response, Eq. \eqref{response}, to periodic forcing, $0.2\cos(\omega t)$ (dashed lines) using the potential $U(x)$ defined in Eq. (\ref{potential_U}) with $\beta=3$ for different values of $\delta=0.1,0.2,0.5$ corresponding to panels (a,b,c) respectively (solid thick black lines) and comparison with the response obtained using two double parabolic potentials with $\beta=3$ and $\beta=\sqrt{3(3-\delta)}$ (solid thin black lines). In the three panels we also show the average of the two solid thin black lines (solid thick grey lines). In all three potentials $\lambda=1$ and $\omega = \pi/50$.}
\label{fig:fig02dis}
\end{figure}

We integrate  Eq. (\ref{eq:lang}) numerically many times using $U(x), U_1(x)$ and $U_2(x)$ for three different values of $\delta$ (0.1,0.2,0.5) and take the average over the trajectories to obtain the response according to Eq. (\ref{response}).  The response for each value of $\delta$ is shown in Fig. \ref{fig:fig02dis} (a)-(c).  Apart from the fluctuations associated with the finite number of trajectories, the response for $U(x)$ is always bounded by the responses for 
 $U_1(x)$ and $U_2(x)$ and is well approximated by their average.  Therefore, the  double parabolic potential reproduces very well the response of the general quartic  of Eq. (\ref{potential_U}) and, for a given $\delta$, provides a confidence interval inside which the response obtained from a quartic potential can be determined.

 
\subsubsection{Neuron activity}

Consider the activity of neurons as a stochastic process with a threshold dynamics, which amplify an input signal.
The membrane voltage, $x(t)$, is well approximated by the following Langevin equation 
\begin{align}
\tau_m\dot{x}(t) = -x(t)+\mu+A\text{cos}(\omega t)+\xi(t),
\end{align}
where $\tau_m$ is a response time-scale, $\mu$ is a drift, $\xi(t)$ is Gaussian white noise and 
the threshold dynamics is represented by resetting the voltage when it reaches $x_{th}$ \cite{plesser1997, shimokawa2000, duarte2008}. 
Namely, the neuron fires when $x(t)$ = $x_{th}$ and the voltage and the phase, $\omega t$, are reset to their values at $t=0$. This situation is analogous to the hopping 
of a Brownian particle from one minimum to another, after which its motion is centered around the new minimum.  Thus, we can 
consider the leaky integrate-and-fire neuron model described above in the double-quadratic potential framework of stochastic resonance described here.

\subsubsection{Manipulating the phase shift in stochastic resonance}

A particularly useful aspect of stochastic resonance is to amplify a weak signal within a background of high-frequency noise.  
In our model the magnitude of amplification is quantified by $\beta^2$ and the phase shift is given by $\phi+\psi$.   Clearly, the 
degree to which the phase shift is favorable or deleterious depends on the nature of the application and hence so too is the 
frequency dependence of the phase shift.  Our approach quantifies these dependencies through Eq. (\ref{eq:sr_mag}) thereby
providing a guide for experiments and simulations.  


\vspace{-0.2 in} 
\subsubsection{Limitations and outlook}

We emphasize that our results treat the most general situation in stochastic resonance. Of particular note is the fact that 
our escape rates in Eq. (\ref{eq:rate}) require neither the adiabatic limit nor that the magnitude of a signal $A$ is asymptotically small. 
However, the complexity of Eq. (\ref{eq:rate}) may pose some challenges for particular applications.   Whereas, assuming that $A \ll 1$, we can use the approximation that 
$\text{exp}[-\frac{A}{\sigma^2}f(t)] \simeq 1-\frac{A}{\sigma^2}f(t)$, thereby showing the amplification of the magnitude 
of a simple trigonometric function $f(t)$. On the other hand, when $A$ is $O(1)$, the exponential form is poorly approximated by a Taylor expansion and amplification of a given signal does not lead to the magnitude of a single frequency component 
in the power spectrum. Furthermore, as discussed above, one must consider the phase shift of the response of a given signal.   Therefore, these issues will constrain the 
use of Fourier transforms in the interpretation and application of the stochastic resonance based  using Eq. (\ref{eq:rate}) across the entire range of parameter space. 


\vspace{-0.20 in}
\section{Conclusion}

We have used asymptotic methods that are central in the theory of differential equations to derive analytical expressions for the entire suite of results in stochastic resonance.  Having previously used this general methodology to analyze the Fokker Planck equation \eqref{eq:mainFP} for a general quartic potential in the non-adiabatic limit \cite{moon2020}, here we have managed to further simplify the problem whilst maintaining the key features of that analysis.  In particular, we approximated the quartic potential of the underlying Ornstein-Uhlenbeck process as two parabolic potentials.  We derived explicit formulae for the escape rates from one basin to the other free from the constraints of the adiabatic limit, and have shown their veracity using direct numerical solutions of the dynamical equations.  Moreover, we have shown numerically that the double-quadratic potential reproduces the results of the continuous quartic potential.  Finally, our results can easily be generalized to multiple frequencies and forcing amplitudes and the ease of use of explicit formulae free a practitioner interested in stochastic resonance from the labor of numerical simulations.

\begin{acknowledgements}
The authors acknowledge the support of Swedish Research Council grant no. 638-2013-9243.  
 \end{acknowledgements}



%

\end{document}